\documentclass[twoside,fleqn]{article}
\usepackage{epsf,espcrc2,amssymb,hyperref}

\def\openone{\leavevmode\hbox{\small1\normalsize\kern-.33em1}}

\def\meff{m_{\rm eff}}

\def\meff{m_{\rm eff}}

\def\a0limit{a \rightarrow 0}

\def\openone{\leavevmode\hbox{\small1\normalsize\kern-.33em1}}

\def\spose#1{\hbox to 0pt{#1\hss}}
\def\ltapprox{\mathrel{\spose{\lower 3pt\hbox{$\mathchar"218$}}
 \raise 2.0pt\hbox{$\mathchar"13C$}}}
\def\gtapprox{\mathrel{\spose{\lower 3pt\hbox{$\mathchar"218$}}
 \raise 2.0pt\hbox{$\mathchar"13E$}}}
\def\inapprox{\mathrel{\spose{\lower 3pt\hbox{$\mathchar"218$}}
 \raise 2.0pt\hbox{$\mathchar"232$}}}

\def\figsizea{3.0}
\def\figsizeb{2.6}

\def\figure#1#2#3#4{\epsfxsize=#4 truein
\centerline{\epsffile{#1}}
\bigskip
\centerline{\vbox{{\bf \noindent Figure #2.} #3}}
\smallskip 
}

\def\one{Eq. \ref{eq:free_loc_cond_3} for $n$ components near $\pi$.
}
\def\two{$\lambda_{\rm tm} = \lambda(\log(-T^2))$ vs. $m_0$ for 
$a_5=1$ and for a topologically non-trivial SU(3) 
background. Diamonds represent 
eigenvectors of $T^2$ with chirality $+1$, pluses with $-1$. 
All eigenvalues are 4-fold degenerate and are indistinguishable.
}

\title{Interacting staggered domain wall fermions}
 
\author{
  Pavlos M.~Vranas \thanks{Speaker. The story in the introduction is from P. Vranas.}
  \address{
    IBM T.J. Watson Research Center, 
    Route 134,
    Yorktown Heights, NY 10598, USA.
  } and  
  George T.~Fleming
  \address{
    Physics Department,
    The Ohio State University,
     Columbus, OH 43210-1168, USA.
  } 
}

\begin{document}

\pagestyle{empty}

\begin{abstract}

The behavior of staggered domain wall fermions in the presence of
gauge fields is presented. In particular, their response to gauge
fields with nontrivial topology is discussed.

\end{abstract}

\maketitle  

\section{Introduction}
\label{sec:introduction}
I remember it very clearly: It was a sunny spring Northern California
afternoon in 1988. I had just passed my Ph.D. exams and was about to
start research. My Ph.D. advisor took his other student (also my
collaborator) and me out to lunch to the place across from the physics
department. This was unusual, so it naturally weighed a lot.  I do not
remember his exact words, but he basically told us about the many
difficulties that would lay ahead should we decide to follow research
in lattice gauge theory.  Now, as if that was not enough, the next
year, at the lattice conference at Capri, the father of this all,
K. Wilson, resigned with a farewell-good-luck-I-am-out-of-here talk.
I am not sure if I heard this directly from K. Wilson or from
anectodal rumors, but whatever might be the case, it was something
like this: `` To be able to do QCD we will need lattice volumes $V
\approx 128^3$, and by the time computers will reach this capability I
will be too old... And there are so many more interesting things in
science to just wait for this...'' Now I see what the great man
meant. Having been involved in building supercomputers for the last
8 years it is my estimate that that volume will be reached around
2012.

Now, these were dire warnings. Needless to say, I did not listen and
the price has been high.  Nevertheless, nearly 14 years later, I am
writing this from a Boston cafe while Lattice 2002 is in progress. But
it is all my advisor's fault...  He introduced 
me to the lattice fermion doubling problem; and it was love 
at first sight...

Where else in theoretical physics can you find a problem that appears
so painfully simple and yet runs so deep? Well, there are a few more
but this is definitely one of them.

And, yes, an extra dimension came in naturally to cater to this
problem. Domain wall fermions (DWF), a revolutionary technique,
were introduced in \cite{Kaplan,Frolov_Slavnov,NN1} (for reviews and 
references see \cite{SDWF_paper1}). And, the extra dimension did not 
come from string theory, nor from any other theory-beyond-the-standard-model,
but from this silly little technical lattice problem.  And I still
feel that we have not yet grasped its full meaning. Because, at the
end of the day, it is the problem of non-perturbative regularization
of chiral gauge theories, which in turn are at the boundary of the
standard model.

And, to add insult to injury, this is one of the main reasons for the
very slow progress in numerical simulations of QCD. The fastest
supercomputers ever built have been traditionally used by
QCD only to feel in their guts of gates and wires the
difficulties of the doubling problem.

What I am trying to say is that the lure is still strong, the problem
is still theoretically very interesting and numerical simulations can
still benefit a great deal from improved lattice fermion methods. So,
for better or for worse, here are staggered domain wall fermions, SDWF
\cite{SDWF_paper1,SDWF_lat01}.

\section{It is not just about doubling}
\label{sec:1}

As is well known, even naive lattice fermions are not equivalent to 16
diagonal flavors. Otherwise there would be 255 naive pions. But there
are only 15. Even in naive lattice fermions there is inherent flavor
mixing.  Traditional Wilson fermions ``hide'' this mixing by raising
the doubler masses but staggered fermions ``retain'' the
mixing. Depending on your point of view this is interesting or plain
annoying or perhaps both.

\section{SDWF}
\label{sec:2}

SDWF are a cross between DWF and staggered fermions. They have an 
exact U(1)$\times$U(1) chiral symmetry for any $L_s$ (where $L_s$ 
is the size of the fifth dimension).
The full SU(4)$\times$SU(4) is recovered at the $L_s \to \infty$ limit.
SDWF should offer an advantage for simulations of the finite temperature 
QCD phase transition.

The SDWF Dirac operator in the Saclay basis \cite{Kluberg-Stern:1983dg} is
given in \cite{SDWF_paper1}. The free theory exhibits localization provided 
that:
\begin{equation}
b^2=
{1 \over 4} \sum_\mu \left[ (1 - \cos{k_\mu}) + (1 - m_0) \right]^2 < 1 
\label{eq:free_loc_cond_3}
\end{equation}
This is shown in Fig. 1 for $n$ components near $\pi$.

\vskip 0.2truein
\figure{m0_range.eps}{1}{\one}{\figsizea}

\noindent
The symmetry content for any $L_s$ is:

\noindent
a) U(1) $\times$ U(1) axial symmetry. The relevant operator is
$(-1)^s \gamma_5 \otimes \xi_5$.

\noindent
b) Rotations by $\pi/2$, in planes perpendicular to the extra dimension 
(as in staggered fermions).

\noindent
c) $\mu$-parity. However, $D_5$ is not invariant unless the $s$ direction 
is also reflected.

\noindent
d) The shift by one lattice spacing is broken for $(m0 - 1/a_5) \ne 0$. 
However, this is not considered a problem for SDWF.

The flavor identification is trickier than staggered \cite{SDWF_paper1}.
The propagator is given in \cite{SDWF_paper1}. The effective
mass $\meff$ in $2n$ dimensions is similar to DWF. For $L_s$ odd:
\begin{equation}
\meff = (1 - {2n \over 4} m_0^2)(m_f + |1 - m_0|^{L_s}) 
\end{equation}
%

\section{The transfer matrix in the Saclay basis}
\label{sec:3}
%

The SDWF transfer matrix is:
\begin{equation}
T = - \left( \begin{array}{cc}
  B^{-1}/a_5        &  B^{-1} C \\
  C^\dagger B^{-1}  & a_5 [C^\dagger B^{-1} C - B]
\end{array} \right) .
\label{eq:transfer_mat}
\end{equation}
One can easily check that:
\begin{equation}[C, \xi_5] = 0, \ \ \{B, \xi_5\} = 0, \ \ \{T, \xi_5\} = 0
\end{equation}
\begin{equation}
B^\dagger = - B \Rightarrow T^\dagger = - T
\end{equation}
T is anti-hermitian.
This is different from DWF. Standard transfer matrix 
manipulations  should be done with the hermitian transfer matrix $T^2$.

\section{Surprise?}
\label{sec:4}
%

The $a_5 \to 0$ Hamiltonian
is proportional to the identity in flavor. No flavor mixing at all...
\begin{equation}
- T^2 = e^{-2 a_5 (H \otimes\openone) }
\end{equation}
\begin{equation}
H = -\left( \begin{array}{cc}
{1 \over 2} \sum_\mu \Delta_\mu + m_0   &  \! \! \! 
C \\
 C^\dagger                              &  \! \! \! 
-{1 \over 2} \sum_\mu \Delta_\mu - m_0
\end{array} \right) 
\end{equation}
This $H$ is very similar to DWF. However, a zero eigenvalue for $H$
does not imply an eigenvalue of magnitude $1$ for $T$ at any $a_5$. Only 
at $a_5 \to 0$. This is different from DWF. For $a_5=0$, all crossings are 
in $0 < m_0 < 2$.

\section{Pseudo - Hamiltonian}
\label{sec:5}
%
To investigate the $m_0$ dependence of $|\lambda(T)|=1$, one can 
eliminate $B^{-1}$ as in DWF. This leads to a pseudo-Hamiltonian 
$H_p$. For $a_5=1$ all crossings are in  $0 < m_0 < 4$.
\begin{equation}
\lambda(H_p) = 0 \Rightarrow \lambda(T) = i \lambda, \ \ \ \lambda = \pm 1 
\end{equation}
\begin{equation}
H_p =  \left( \begin{array}{cc}
    1 + a_5 \lambda i B   &  a_5 C \\
    a_5 C^\dagger        & -1 - a_5 \lambda i B
\end{array} \right)
\end{equation}
%

\section{The spectrum of T}
\label{sec:6}
The spectrum of $T$ is doubly degenerate because $\{T, \xi_5\}=0$.
For a degenerate four-flavor theory, $log(-T^2)$ must have four
zero crossings at the same $m_0$ with the same chirality.
For {\it any} SU(2) field the degeneracy is always four-fold.

For a smooth non-trivial SU(3) instanton configuration (plaquette $= 0.05$) 
the eigenvalues are almost exactly four-fold degenerate. This can be seen 
in Fig. 2. At every crossing, four eigenvalues of the {\cal same} chirality 
cross. For a very rough SU(3) gauge field configuration (plaquette $= 0.85$ ) 
the four-fold degeneracy of $\log(-T^2)$ splits to two-fold, but only by 
a small amount. This is shown in Table 1 for one of the worst cases 
on a $2^4$ lattice.

\figure{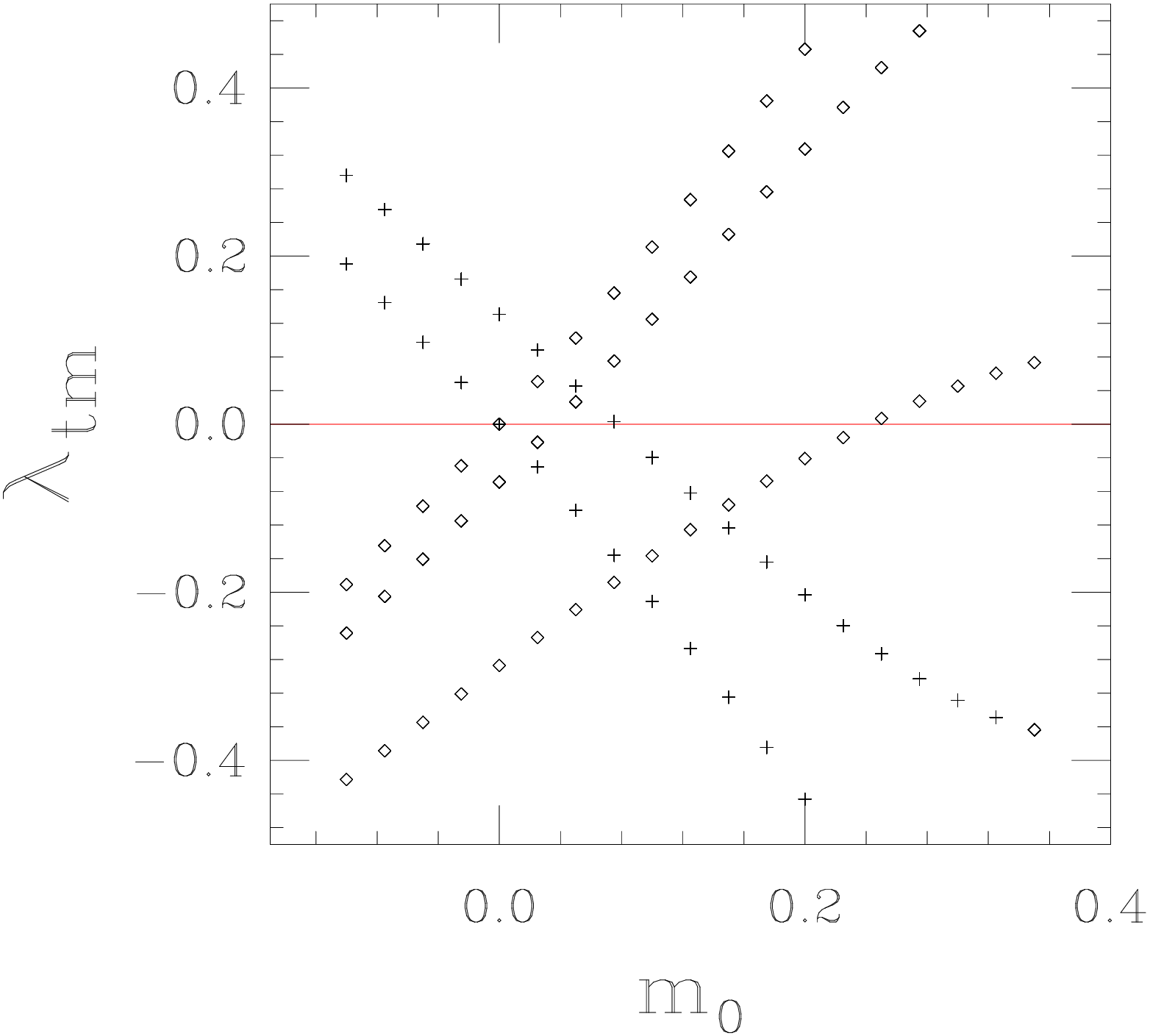}{2}{\two}{\figsizeb}
%

\section{Going back to where we came from ...}
\label{sec:7}
Numerical simulations are done in the single component basis. For
various approaches see \cite{SDWF_paper1}. For example, the standard
transcription from the Saclay basis to the single component basis can
be used.  When gauge fields are present, the transcription carries a
Jacobian that is a function of the gauge fields. Since the gauge field
dependent part of the fermionic and PV actions is identical, the
Jacobians must cancel (for details see \cite{SDWF_paper1}).

\begin{table}
\begin{tabular}{||l|l||l|l|}        \hline
$m_0$   & $log{\lambda(-T^2)}$ & $m_0$  & $log{\lambda(-T^2)}$ \\ \hline
0.3	&	-0.262638      &  0.3	&	0.0260566    \\ \hline
0.3	&	-0.262638      &  0.3	&	0.0260566    \\ \hline
0.3	&	-0.261551      &  0.3	&	0.0272779    \\ \hline
0.3	&	-0.261551      &  0.3	&	0.0272779    \\ \hline
\end{tabular}
\caption{The near zero spectrum of $log(-T^2)$ for $m_0=0.3$.}
\end{table}

\section{Still....}
\label{sec:8}

\noindent
1) For QCD the nearly four-fold crossing degeneracy must be investigated 
more thoroughly.

\noindent
2) SDWF in the single component basis must be tested.

\noindent
3) A full Hamiltonian analysis is needed.

\noindent
4) The Kogut-Sinclair \cite{Kogut:1997mj}
4-fermion interaction with SDWF can be used to span the finite temperature 
QCD phase transition at zero quark mass.

\noindent
5) Is there something new that SDWF have
revealed about the inherent lattice fermion flavor mixing?

\section*{Acknowledgments}
Our special thanks to Prof. J.B. Kogut for his continued encouragement 
and support.


\end{document}